\documentclass[%
reprint,
superscriptaddress,
preprintnumbers,
 amsmath,amssymb,
aps,
prc,
]{revtex4-2}
\usepackage{amssymb}
\usepackage{graphicx}% Include figure files
\usepackage{dcolumn}% Align table columns on decimal point
\usepackage{bm}% bold math
\usepackage{hyperref}% add hypertext capabilities

\begin{document}

\preprint{Published in \href{http://www.ptep-online.com/2017/PP-50-02.PDF}{Prog. Phys. 13, 150 (2017)}}

\title{Configuration mixing in particle decay and reaction}

\author{Elsayed K. Elmaghraby}
\email[Institutional email: ]{elsayed.elmaghraby@eaea.org.eg}
\email[Alt. email: ]{e.m.k.elmaghraby@gmail.com}
 \affiliation{Experimental Nuclear Physics Department, Nuclear Research Centre, Egyptian Atomic Energy Authority, Cairo 13759, Egypt.}

\date{Jan. 2016}% It is always \today, today,
 % but any date may be explicitly specified

\begin{abstract}
\begin{description}
\item[Background] Decay constants of unstable nuclei and particles are quantum constants. Recent controversy on the existence (versus non-existence) of variability in the observation of decay rate can be settled by considering mixing in decay configuration.
\item[Purpose]
Variability in decay rate was investigated based on the available information of beta decay rate data, solar neutrino flux, and energy distribution. 
\item[Method]
Full systematic analysis of the oscillatory behavior was carried out. Based on the zero threshold energy for neutrino absorption in beta emitters, a model for configuration mixing between two distinct beta disintegration modes $\beta^\nu$-disintegration (electron from neutrino interaction) and the $\beta^-$-disintegration (electron from natural decay) was proposed.
\item[Results]
The phenomenon of variability in beta decay rate was related to the possible exothermic neutrino absorption by unstable nuclei which, in principle, should include the whole range of flux energies involving flux with energy below the $^{71}$Ga threshold at 0.23 MeV. These two disintegration modes occur independently and model for their apparent mixing rate was proposed.
\item[Conclusions]
The configuration mixing between the two modes cause depletion of radioactive nuclei which is subject to change with seasonal solar neutrino variability. Ability to detect this variability was found to be dependent on the Q-value of the $\beta^\nu$ disintegration and detection instrument setup. Value of neutrino cross section weighted by the ratio between $\beta^\nu$ and $\beta^-$ detection efficiencies was found to be in the range $10^{-44}$ to $10^{-36}$ cm$^{2}$. For experiments that uses the end point to determine the neutrino mass, interference due to mixing should be taken into account.
\end{description}
\end{abstract}%
\maketitle
\clearpage
\section{Introduction}

Anomalous behavior in radioisotopes activity  was reported by several scientists, they considered it as influence of solar proximity and activity. Several scientist are in favor of the influence of solar activity/distance on the decay rate. Early results of Alburger et al. \cite{Alburger1986168} are based on normalizing the count rate ratio of $^{32}$Si/$^{32}$P decay rate. Siegert et al. \cite{Siegert19981397} had reported oscillatory behavior of $^{226}$Ra, $^{152}$Eu, and $^{154}$Eu. Jenkins et al. \cite{Sturrock20168} had studied these cases and reported several new data and measurements. Most investigator had reported seasonal relation between oscillatory behavior and the earth's position with respect to its sun's orbit; referring to the neutrino influence to the decay process.

Several other scientists oppose the connection between sun and the phenomenon. In one of the oppose thoughts, scientists may consider the rare neutrino events in experiments like Ice Cube and Sudbury Neutrino Observatory \cite{Aharmim2007045502}; yet, the energy threshold of there detection system may not fall below $^{71}$Ga border at 0.233 MeV (3.5-5 MeV for electron scattering\cite{Bellini2010033006}, 1.44 MeV for $d(\nu_e,e)pp$ interaction.  In all measurements no relation between half-life and the existence of this phenomenon was reported. Several other oppose reports, based on measurements by different techniques, were published, see Refs. \cite{Cooper2009267,Bellotti2013116,Alexeyev201323,Pomme2016281,Bikit201338}.

In the present work, full systematic analysis and treatment  of the oscillatory behavior was performed in order to reconcile these viewpoints. Based on the zero threshold energy for neutrino absorption beta emitters, a model for configuration mixing between distinct $\beta^\nu$-disintegration (the electrons from neutrino interaction) and the $\beta^-$-disintegration (the electrons from natural decay) was proposed.

\section{Model for analysis}

The majority of solar neutrino are with electron flavor associated with proton burn-up processes ($\phi_{\nu_e,pp}$=6$\pm$0.8$\times 10^{^{10}}$ cm$^{-1}$s$^{-1}$) with maximum energy around 0.41 MeV\cite{Abdurashitov2009015807}. During solar flares protons stimulates production of  pions ~ / ~muons; $\pi^+$ ($\pi^-$) decays into $\nu_\mu$($\overline{\nu}_\mu$) with $\mu^+$ ($\mu^-$), later partners decay and emit $\nu_e$ ($\overline{\nu}_e$) together with $\overline{\nu}_\mu$ ($\nu_\mu$)\cite{Ryazhskaya2002358} total flux is of order $10^{9}$ cm$^{-2}$s$^{-1}$ and has energy up to 10 MeV.

Rare reaction of neutrino with stable isotopes is attributed to its small coupling with $W^{\pm}$ and $Z^0$ bosons, and higher threshold of reaction kinematics. Coupling with $Z^0$ may be not appreciated due to non-existence of
flavor changing neutral currents. If happened, an electron neutrino in the vicinity of the nucleus couples with a W boson emitting a $\beta^\nu$ and induces beta transformation in the nucleus. Threshold energy of neutrino capture in $^{37}$Cl is about 0.813 MeV compared to 0.233  MeV in $^{61}$Ga, these isotopes are used as monitor for $^8$Be neutrinos. Radioactive isotopes, on the other hand, have excess energy to deliver due to positive $Q$-values as illustrated in Table \ref{Tab.1}. Hence, one can conclude that the solar influence on the \emph{apparent} decay rate is associated mixing of specific mode of \emph{disintegration} in consequence of neutrino capture in nuclei with the natural disintegration rate. The apparent decay rate of radioactive isotopes, $\lambda^\prime$ may be split into two terms; a term for usual disintegration of the nucleus labelled $\lambda_d$ and a terms for neutrino interaction. Presumably, neutral current will contribute to scattering only. $\beta^-$-decay rate is proportional to the  matrix element of the decay, $|M_d|^2$ while the reaction terms are associated with matrix elements of neutrino interaction with charged current, $|M_{\nu W^{\pm}}|^2$.
\begin{equation}\label{Eq.ActingFunction0}
  \begin{split}
  N(t)\lambda^\prime=&N(t)\lambda_d+
  N(t) \sum_{\text{flavors}} \phi_{\nu}{(t)}N_N\left<K_{\beta^\nu}(Q)\right>\sigma_{\nu n}
  \end{split}
\end{equation}
\noindent The summation is taken over all possible neutrino flavors. Here, N(t) is the number of nuclei at time t,  $\left<K(Q)\right>$ is the factor representing the modification of nucleon properties in the nuclear medium, which can be investigated by nucleon induced nuclear reactions\cite{Elmaghraby2009c,Elmaghraby2008}; $\left<K(Q)\right>$ depends on the $Q$-value of the reaction and the state of the nucleus upon interaction. The in-medium neutrino cross section $\sigma_{\nu}$ can replace  $\left<K(Q)\right>$ $\sigma_{\nu n}$.

\begin{figure}
  \centering
  % Requires \usepackage{graphicx}
  \includegraphics[width=8cm]{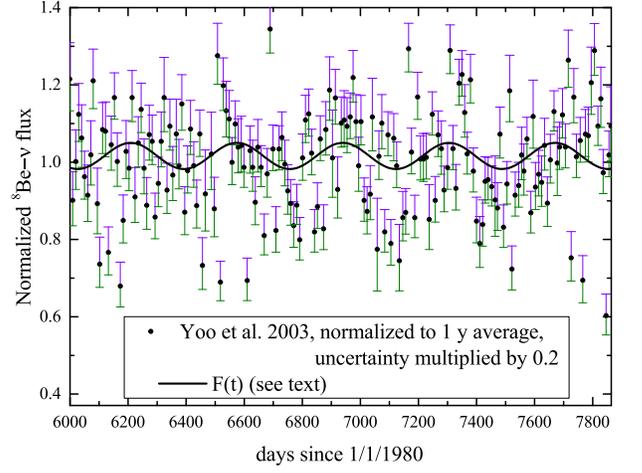}
  \caption{Normalized measurements of $^8$Be neutrino variation by Yoo et al. \cite{Yoo2003092002} in comparison with predictions of $F(t)$ function in Eq. \ref{Eq.TimeVa}. }\label{fig0}
\end{figure}

The flux would be altered with the change in earth to sun distance $R$. Hence the time varying function is inversely proportional to the area of a sphere centered at sun. The radius vector has the form
\begin{equation}\label{Eq.RadiusVector1}
  R= a \frac{1-\epsilon^2}{1+\epsilon \cos(\theta)}, \;\; \theta\approx\omega t
\end{equation}
Where $\epsilon$ is the eccentricity of earth's orbit (now, 0.0167\cite{Simon1994663S}) and the cosine argument is the angle relative to the distance of closest approach (2-4 January) in which value equals to $R=a(1-\epsilon)$. $\omega=2\pi/T_\omega$ is the average orbital velocity, and $T_\omega$ is the duration of earth's years in days. The approximate sign is introduced because earth's spend much more time at larger distance from the sun than in the near distances.  Assuming that the average flux  ($\phi_\nu^{(0)}$) occurs at time $t_0$ during the revolution around the sun, the flux at any other time will be
%\begin{equation}\label{Eq:AverageFluxandFlux}  \phi_\nu=\phi_\nu^{(0)}(a \frac{1-\epsilon^2}{1+\epsilon \cos \omega t_0})^2 /(a\frac{1-\epsilon^2}{1+\epsilon \cos \omega t})^2\end{equation}
\begin{equation}\label{Eq:AverageFluxandFlux}
  \phi_\nu(t)=\phi_\nu^{(0)}
F(t), %\quad F(t)=\frac{(1+\epsilon \cos(\omega t))^2}{(1+\epsilon \cos (\omega t_0))^2}
\end{equation}
\begin{equation}\label{Eq.TimeVa}
F(t)=\frac{(1+\epsilon\cos(\omega t))^2}{(1+\epsilon \cos (\omega t_0))^2}
\end{equation}
Here, $\phi_{\nu}^{(0)}$ is the average flux of neutrinos reaching earth's surface (about $6.65\times10^{10}$ cm$^{-2}$s$^{-1}$ as average of all sun's producing routes\cite{soler2008neutrinos}, in which only 2.3$\times 10^{6}$ cm$^{-2}$s$^{-1}$ are from $^8$Be. Comparison between $F(t)$ (taking $t_0=0$) and normalized seasonal variation of $^8$Be neutrons (data taken from Yoo et al. \cite{Yoo2003092002} and normalised to its yearly average) is represented in Fig. \ref{fig0}. $F(t)$ gives the averaged trend of Yoo et al. data within the experimental uncertainty of measurement.

For simplicity,  and due to nature of available data of being related to oscillatory behavior, effect of cosmological neutrinos will be disregarded.
Additionally, non-predominant radioactive isotopes should have the neutrino-induced beta disintegration of contribution  much smaller than that of the $\beta^-$-decay; hence, $\lambda_d$ can be replaced by the laboratory decay constant, $\lambda$, with good precision. The apparent decay rate for specific interaction current can be described by the formulae
%$Z^0$ components will have large contribution due to its relation with neutrino scattering; it may transfer energy to nucleus which may change route of decay.
\begin{equation}\label{Eq.ActingFunction}
  \begin{split}
\lambda^\prime &\Rightarrow\lambda+\phi_{\nu}^{(0)}N_N\sigma_{\nu}F(t)\\
%&\equiv \lambda(1+\frac{\phi_\nu^{(0)}\Gamma_\mu}{\lambda}\sin(\omega t + \delta))\\
  \end{split}
\end{equation}
Where, $\phi_{\nu}^{(0)}$, $N_N$, and $\sigma_{\nu}=$ $\left<K\right>$$\sigma_{\nu n}$ are related to the considered current and the disintegrated nucleus.
Differential nuclear decay rate is simply described by the rate equation $dN(t)/dt=-N(t)\lambda^\prime$. Upon integration, the number of survived nuclei become
\begin{equation}\label{Eq.Nt}
  \begin{split}
  N(t)=&N(0) \exp\left(-\lambda t -\mu\left(1+\frac{\epsilon^2 }{2}\right)t\right) \times \\ & \exp\left(-2\epsilon\mu \left(1+\frac{\epsilon}{4} \cos\left(\omega t\right)\right)\sin\left(\omega t\right)\right),\\
  \mu=&\frac{\phi_{\nu}^{(0)}N_N \sigma_{\nu} }{\omega (1+\epsilon \cos (\omega t_0))^2}
  \end{split}
\end{equation}

\noindent The first exponential represent the \emph{depletion} of nuclei with neutrino interaction together with the radioactive decay.
%During the period of short-term measurements the dilation term goes to unity; otherwise, it must be taken into consideration.
The second exponential can be represented as
\begin{equation}\label{Eq.Nt22}
  1+\sum_{i=1}\frac{(-1)^i}{i!}\left(2\epsilon\mu \left(1+\frac{\epsilon}{4} \cos\left(\omega t\right)\right)\sin\left(\omega t\right)\right)^i
\end{equation}
\noindent The value of $2\epsilon\mu<<1$; hence, only the first term in the summation is effective. I.e.,
\begin{equation}\label{Eq.Nt2}
  \begin{split}
  N(t)=& e^{-\lambda t -\mu\left(1+\frac{\epsilon^2 }{2}\right)t} \left( N(0)-A\left(1+\frac{\epsilon}{4} \cos\left(\omega t\right)\right)\sin\left(\omega t\right)\right)
  \end{split}
\end{equation}
Which reveal seasonal variability. The amplitude of the oscillation is $A\,=\,2\,N(0)\, \epsilon \, \mu$ with the depletion factor $\exp\Big(-\lambda t\Big.$ $\Big.-\mu\left(1+\frac{\epsilon^2 }{2}\right)t\Big)$; depletion factor reaches unity for long-lived isotopes with relative short-term measurements.

The method of normalization of data, mentioned in context, is intended to remove the effect of isotope decay rate and give the residual of neutrino interaction. So, when normalized to 1, the normalized fraction (proportional to decay rate or detector count) becomes
\begin{equation}\label{Eq.NtNor}
  \begin{split}
  \overline{N(t)}=&  \left(1-A e^{-\lambda t  } e^{-\mu\left(1+\frac{\epsilon^2 }{2}\right)t}\left(1+\frac{\epsilon}{4} \cos\left(\omega t\right)\right)\sin\left(\omega t\right)\right)
  \end{split}
\end{equation}
\noindent Similarly, for normalization of the ratio between two isotope 1 and 2,
\begin{equation}\label{Eq.NtNor2}
  \begin{split}
  \overline{N_{1/2}(t)}\approx& 1- 2 \frac{N_1(0)}{N_2(0)}e^{-(\lambda_1-\lambda_2) t  } e^{-(\mu_1-\mu_2)\left(1+\frac{\epsilon^2 }{2}\right)t} \times\\&  \frac{\epsilon\mu_1\left(1+\frac{\epsilon}{4} \cos(\omega t)\right)\sin(\omega t)}{1-2\epsilon\mu_2\left(1+\frac{\epsilon}{4} \cos(\omega t)\right)\sin(\omega t)}
  \end{split}
\end{equation}
\noindent Which is not a complete sinusoidal variation.  The amplitude and depletion factors in case of two activity ratio becomes
\begin{equation}\label{Eq.NtNor2amp}
  \begin{split}
A_{1/2}=2 \frac{N_1(0)}{N_2(0)}\frac{\epsilon\mu_1}{1-2\epsilon\mu_2}e^{-(\lambda_1-\lambda_2) t  } e^{-(\mu_1-\mu_2)\left(1+\frac{\epsilon^2 }{2}\right)t}
  \end{split}
\end{equation}
This depletion term can be ignored if both isotopes have comparable half-life and mass.

\section{Discussion}

Normalized oscillatory data, were collected for the decay of isotopes given in Table \ref{Tab.1}.  Because we need to have a starting point, data retrieved relative to 1 Jan. 1980. The time shift, $t_0$, was obtained using least square fitting of every data set with Eq. \ref{Eq.NtNor} by shifting time with free parameter--say $t_1$. Results are illustrated in Table \ref{Tab.1} in which a shift of -120$\pm 14$(1$\sigma$-stat.)$\pm$5(1$\sigma$-syst.) days was found; i.e. the average flux received on earth from the sun occurs around end of October (or, alternatively, May first.) This is consistent with data given in measurement of $^8$Be neutrino variation by Yoo et al. \cite{Yoo2003092002}.

\begin{table*}
  \centering
  \caption{Data of seasonal variability of radioactive disintegration. Unit of $\xi \sigma_{\nu}$ is cm$^{2}$, Q-value is calculated from AME2003 atomic mass evaluation \cite{Audi2003-2} in the unit of MeV; $t_1$ is the time shift in days.}\label{Tab.1}
\begin{tabular}{c|c|c|c|c|c|c}
\hline\hline

   Isotope &        Ref & A $\times10^{+4}$            &     $-t_1$ &       $\xi\sigma_{\nu}\times10^{41}$ &                  $Q$ &        $E_{th}$  \\
\hline

%$^{37}$Cl &            &                                  &              &            &      -0.814 &      0.814 \\

%$^{71}$Ga &            &     &      & &     -0.233 &      0.233 \\ \hline
    $^3$H & \cite{Kossert2015172,Pomme2017beta} &      5.29 $\pm$2$\pm$6 &        -10$\pm$30 &    1.8$\pm$0.4$\pm$2 $\times10^{-6}$ &     0.0186 &          0 \\

    $^3$H & \cite{Falkenberg200132} &       38.4$\pm$0.8$\pm$35 &        138$\pm$1 &    13$\pm$0.1$\pm$12$\times10^{-6}$ &     0.0186 &          0 \\

$^{32}$Si & \cite{Alburger1986168} &       10.8$\pm$2$\pm$5 &        109$\pm$12 &    1.15$\pm$0.01$\pm$0.5$\times10^{-3}$ &     0.2243 &          0 \\

$^{32}$Si/$^{36}$Cl & \cite{Alburger1986168,Jenkins200942,JavorsekII2010173} &       15.8$\pm$0.67$\pm$7 &        126$\pm$5 &      &         &            \\

$^{36}$Cl & \cite{Jenkins201281} &         19$\pm$0.9$\pm$10 &        160$\pm$5 &    4.52$\pm$0.01$\pm$2.4 &     0.7097 &          0 \\

$^{152}$Eu & \cite{Schrader2007S53,Schrader20101583,Sturrock201442} &      8.4$\pm$0.3$\pm$2 &        113$\pm$3 &    3.51$\pm$0.01$\pm$0.79 &     1.8197 &          0 \\

$^{154}$Eu & \cite{Schrader2007S53,Schrader20101583,Sturrock201442} &      8.5$\pm$0.4$\pm$3 &        121$\pm$2 &    8.7$\pm$0.004$\pm$3 &     1.9688 &          0 \\

%121mSn/241Am & \cite{Rech2002057302} &          2$\pm$  0.5 $\pm$ 0.6 &        120     &    233$\pm$0.820582$\pm$ 69.9136 &      0.391 &          0 \\

$^{214}$Bi& \cite{Stancil2017385} &         31$\pm$2$\pm$17 &        119$\pm$6 &    39.7$\pm$0.02$\pm$22 &     3.2701 &          0 \\

$^{214}$Bi& \cite{Stancil2017385}  &         30$\pm$2$\pm$12 &        118$\pm$5 &    38.5$\pm$0.02$\pm$15 &     3.2701 &          0 \\

\hline

% $^{22}$Na/$^{44}$Ti$^{c}$ & \cite{Keefe2013297,Norman1998} & 4.3$\pm$0.9$\pm$3 &                    &            &            &            \\

$^{85}$Kr & \cite{Schrader20101583,Sturrock201442} &       7.2$\pm$0.35$\pm$1.5 &        113$\pm$3 &              &      0.687 &          0 \\

$^{90}$Sr & \cite{Schrader20101583,Sturrock201442} &      8.8$\pm$0.4$\pm$2 &        121$\pm$3 &             &      0.546 &          0 \\

$^{108}$Ag & \cite{Schrader20101583,Sturrock201442} &        8.6$\pm$0.3$\pm$2 &        126$\pm$2 &                    &       1.76 &          0 \\

$^{133}$Ba & \cite{Schrader20101583,Sturrock201442} &      6.18$\pm$0.6$\pm$4 &        119$\pm$5 &        &     -2.061 &      2.061 \\

\hline
$^{226}$Ra & \cite{Jenkins200942,JavorsekII2010173} &      10.1$\pm$0.3$\pm$3 &        105$\pm$20 &    83$\pm$0.02$\pm$20$\times10^{-3}$ &     diverse &      \\

$^{226}$Ra & \cite{Siegert19981397} &      11.9$\pm$0.2$\pm$2 &        125$\pm$2 &    99$\pm$0.01$\pm$20$\times10^{-3}$ &    diverse &     \\

%$^{226}$Ra & \cite{Siegert19981397,Jenkins200942,JavorsekII2010173} &       7.82$\pm$0.2$\pm$3 &         99$\pm$20 &    65$\pm$0.01$\pm$20$\times10^{-3}$ &     diverse &       \\

\hline
   \end{tabular}
\end{table*}

Before going further in the discussion, we must apprehend measurement techniques and circumstance of each experiment. The correlation between earth sun distance and decay rate for $^{32}$Si and $^{226}$Ra was reported by Jenkins et al. \cite{Jenkins200942} based on Alburger et al. \cite{Alburger1986168} and Siegert et al. \cite{Siegert19981397}; those measurements are based on the $\beta$ spectrum measurements. Alburger and coworkers used end-window gas-flow proportional counter system and a liquid/plastic scintillation detectors and Siegert and coworkers used both $4\pi$ ionization chamber and Ge and Si
semiconductor detectors with reference to ionization chamber measurements. Same group of Ref. \cite{Jenkins200942} and  others in later work \cite{Mohsinally201629} had measured the $^{54}$Mn using the 834.8 keV$\gamma$-line during 2 years without significant seasonal variation, they only report a connection with solar storm. Similar results appeared after solar flare\cite{Jenkins2009407}. Variation of $^{36}$Cl decay rate was reported by BNL group\cite{Jenkins201281} using Geiger-M\"{u}ller counter and in PTB-2014 measurements\cite{Kossert201433} using the triple-to-double coincidence ratio liquid scintillation counting system. PTB-2014 detection system excluded the idea of time varying decay rate while the BNL measurements prove the phenomenon. Power spectrum analysis \cite{Keefe2013297,Jenkins201281,Sturrock2010251,Jenkins201350,JavorsekII2010173,Sturrock2012755} reveal several spectral frequencies especially at 1 y$^{-1}$. Some explanations of seasonal variation of decay rate were related to decoherence in gravitational field \cite{Singleton2015941} and internal sun modes\cite{Sturrock201362}. An experiment was performed for $^{222}$Rn decay in controlled environment showed dependence on the angular emission of gamma ray \cite{Steinitz2015731} and daily behavior \cite{Jenkins2010332,Bellotti2015526,Sturrock20168}; however Bellotti et al.\cite{Bellotti2015526} excluded the sun influenced decay rate in support with there earlier work \cite{Bellotti2013116}.  Ware et al. \cite{Ware2015073505} returned the variation to change in the pressure of counting chamber during the seasonal variation.

Opposition to the connection between sun's and the variability phenomenon of apparent decay rate came out as a consequence of measurements, as well.  No significant deviations from exponential decay are observed in Cassini spacecraft power production due to the decay of $^{238}$Pu\cite{Cooper2009267}. Bellotti et al.\cite{Bellotti2013116} studied decay of $^{40}$K, $^{137}$Cs and $^{232}$Th using NaI and Ge detectors with no significant effect of earth-sun distance.  Same results had been reported by Alexeyev et al.\cite{Alexeyev201323} in the alpha decay of $^{214}$Po measured by $\alpha$-particle absorption. However, Stancil et al. \cite{Stancil2017385} detected seasonal variation in the gamma transition in $^{214}$Po due to $^{214}$Bi decay in radium chain. Others \cite{Alexeyev2016986} had reported seasonal variation in life time of  $^{214}$Po. Recently, Pomm\'{e} et al. \cite{Pomme2016281} re-performed measurements in several laboratories by all possible measurement techniques including ionisation chamber, HPGe detector, silicon detector, proportional counter, anti-coincidence counting, triple-to-double coincidence, liquid scintillation, CsI(Tl) spectrometer, internal gas counting. They returned the phenomenon to lower stability of instruments. Bikit et al \cite{Bikit201338} investigated the $^3$H decay rate by measured by liquid scintillation and related the fluctuation of  the high-energy tail of the beta spectrum to instrumental instability.

The techniques of measurements is different among these two parties.
{Among all measurements given above, all techniques that are based on detecting $\beta$-radiation, or combined $\beta$- $\gamma$-radiation coming from its daughter, had signaled  variability. Which can be explained as a consequence of the mixing between $\beta^\nu$ and $\beta^{-}$ disintegrations. In such case, both terms in Eq. \ref{Eq.ActingFunction} are effective and the apparent decay rate should be influenced by solar proximity and activity. On the other hand, techniques that uses specific decay parameter such as specific $\gamma$-line from $\beta^{\pm}$- or $\alpha$-decay may not be able to recorded any variability because the oscillatory part of configuration mixing in Eq. \ref{Eq.ActingFunction} is not operative. With pure $\alpha$-emitters like $^{241}$Am and $^{226}$Ra, the mixing oscillatory term will change sign and time shift of half-period may appear. In accordance to Siegert et al. \cite{Siegert19981397} results, time shift of a half period in the fluctuation measured between $4\pi \gamma$-ionization chamber measurement of $^{226}$Ra and measurements of $^{152,154}$Eu by GeLi semiconductor detectors was found.
Hence, both parties judged existence or non-existence of the phenomenon based on their technique of measuring it. Each team draw the correct picture of his viewpoint; that is determined by whether the mixing part of Eqs. \ref{Eq.ActingFunction} and \ref{Eq.NtNor} were taken into account or not.}

The $\beta^{\nu}$ energy spectrum should, in principle, reflects the energy distribution of neutrino and the structure of residual nucleus. In $\beta^\nu$-decay, all energy of neutrino plus the major contribution of mass access (Q-value) is transferred to the beta particle. The higher the Q-value, the higher the energy of the emitted $\beta^\nu$. This is another source of disagreement among both teams supporting and declining the phenomenon. Observation of the phenomenon is determined by the ability of their system to detect $\beta^\nu$ or the specific $\gamma$-transition or mass loss subsequent the disintegration.  Detection volume, in general, is selective to a band of radiation energy. Ionization chamber detects gamma radiation and fraction of beta radiation above few hundreds eV\cite{Schrader2007S53}. Additionally, higher energy of $\beta^\nu$ have higher value of detection efficiency. Counting of $\beta^\nu$, and $\beta^-$, and/or their corresponding $\gamma$-ray from nuclei, have different efficiencies due to difference in energy distribution and end-point(c.f. \cite{Schrader2000325}); literally, $\beta^\nu$ has no end-point. Hence, each count rate must be related to its efficiency; i.e. the amplitude of the variation must be modified by a ratio--say $\xi$--between $\beta^\nu$ counting efficiency and $\beta^-$ counting efficiency; which depends on the $\beta^\nu$ energy and the measurement setup.  If variation occurs, it would be reflected on the counting rate. The value of $\xi \sigma_{\nu}$ represent an weighted cross section and it was calculated as a Whole in Table \ref{Tab.1}.

\begin{figure}[ht]
  \centering
  % Requires \usepackage{graphicx}
   \includegraphics[width=8cm]{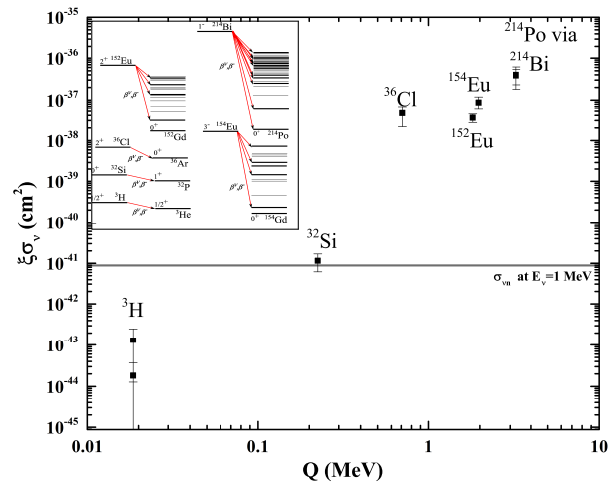}\\
  \caption{Value of the reduced cross section $\xi\sigma_{\nu}$ in the unit of cm$^{2}$ in correlation with the $Q$-value of the possible $\beta^\nu$-disintegration. Line represent the value of $\sigma_{\nu}$=0.881$\times$$10^{-38}$ $E_\nu(GeV)$  cm$^{2}$ at $E_\nu$=1 MeV. Insert: possible disintegration probabilities of represented isotopes to levels in daughter nuclei. }\label{fig1}
\end{figure}

The amplitude of the variability was obtained from each dataset by fitting using Eq. \ref{Eq.NtNor}; results are represented in Table \ref{Tab.1}. The value of  $N(0)$ (alternatively, mass or activity) was found for $^3$H (assuming 1-20 g of $^3$H$_2$O as for PTB measurements catalogue of activity standards\cite{NIST2017C}), $^{32}$Si (0.0477 g of $^{32}$SiO$_2$\cite{Alburger1986168}), $^{36}$Cl(0.4 $\mu$Ci\cite{Kossert201433}), $^{152}$Eu (40 MBq \cite{Schrader2007S53,Schrader20101583,Sturrock201442}), $^{154}$Eu (2.5 MBq \cite{Schrader2007S53,Schrader20101583,Sturrock201442}), and $^{214}$Bi(2 $\mu$Ci\cite{Stancil2017385}), see Table \ref{Tab.1}. Mass, activity, and/or number of decaying atoms were not reported for other datasets. Then, the value ($\xi\sigma_{\nu}$) are calculated only for the said isotopes. A plot for the  variation of $\xi\sigma_{\nu}$ with $Q$-value is represented in Fig. \ref{fig1}. The known limit of $\nu_e$-neutron cross section is $\sigma_{\nu}$= 0.881$\times$$10^{-38}$ $E_\nu(GeV)$  cm$^{2}$ which is represented by the line in Fig. \ref{fig1} for electron neutrino with $E_\nu$=1 MeV considering $\xi=1$. The increase of $\xi\sigma_{\nu}$ with $Q$-value confirms the mentioned hypothesis of existence of instrumental setting participation in the detection of the variability of apparent decay rate.

%An effect of the Q-value may manifest itself on the value of $\langle K(Q)\rangle$. Until now there is no complete theoretical framework for neutrino interaction inside nuclear medium.

In the insert of Fig. \ref{fig1}, decay schemes of said isotopes is represented. The $\beta^\nu$ spectrum is expected to have definite spectrum corresponding to direct transition to levels in daughter nuclei in similarity to neutrinoless double beta decay; one of the possible broadening that could occur is due to original energy distribution of neutrinos. Sensitive detector like KATRINE\cite{MERTENS2015267} can be used to detect such energy distribution in $^{3}$H; fortunately,  neutrinoless double beta decay cannot occur in case of $^{3}$H without fission of the whole nucleus. Disintegration of $^{3}$H, $^{32}$Si and $^{36}$Cl have single possible transition for both $\beta^\nu$ and $\beta^-$ decays. The maximum energy of $\beta^\nu$-$^{3}$H decay is expected to be 0.42 MeV with $\xi\sigma_{\nu}$=1.82$\pm$0.4 stat. $\pm$ 2 syst. $\times10^{-44}$ cm$^{2}$ as calculated from Pomme et al. \cite{Pomme2017beta} data, and 13.2$\pm$0.1 stat. $\pm$ 11 syst. $\times10^{-44}$ cm$^{2}$ as calculated from Falkenberg \cite{Falkenberg200132} data. Systematic uncertainties are mostly related to unknown mass of the material. The BNL data of $^{226}$Ra and other data of radium had been evaluated but was not represented in Fig. \ref{fig1}. $^{226}$Ra has threshold  for $\beta^\nu$ decay of 0.641 MeV, its daughters have possible beta decay probability, that is why variability can be observed \cite{Schrader2007S53,Schrader20101583,Sturrock201442,Jenkins200942,JavorsekII2010173,Siegert19981397}.
The phenomenon disappeared when $\alpha$-detection system is used \cite{Pomme2016alpha}.

%Decay in chain introduces delay in response to decay associated with the half-lives of its isotopes\cite{Elmaghraby???}.

%Several sinusodal fitting paramerers are given in litratures \cite{Mueterthies2016arXiv160703541M}. However, from the present fitting data of each dataset, given in Table \ref, the value of $\sigma_{\nu}$ had been deduced.

\section{Conclusion}
Rare mixed configuration between neutrino induced beta disintegration and natural beta disintegration may exists. These two distinct classes of beta decay could, in principle, explain the variation of apparent decay rate of radioactive isotopes with sun proximity. The circumstances of detection and instrumental ability determine whether to detect pure natural disintegration or the mixed mode. Configuration mixing between $\beta^\nu$ and $\beta^-$ is, presumably, happen among all existing $\beta^-$ emitters. The mixing in configuration of decay and reaction can be extended to all particles and nuclei. It must be taken into account in the in high precision measurements of neutrino mass. Mixing may be of significance for nucleosynthesis in astronomical object.

%\bibliography{Exciton}
%\bibliographystyle{e-journal}
%apsrev4-2.bst 2019-01-14 (MD) hand-edited version of apsrev4-1.bst
%Control: key (0)
%Control: author (72) initials jnrlst
%Control: editor formatted (1) identically to author
%Control: production of article title (-1) disabled
%Control: page (0) single
%Control: year (1) truncated
%Control: production of eprint (0) enabled
%

\end{document}